\documentstyle[prl,aps,twocolumn,psfig,epsf]{revtex}

\begin{document}
\title{Enhancement  of the Josephson current by an exchange field in
superconductor-ferromagnet structures }
\author{F. S. Bergeret $^{1 }$, A. F. Volkov$^{1,2}$ and K. B. Efetov$^{1,3}$}
\address{$^{(1)}$Theoretische Physik III,\\
Ruhr-Universit\"{a}t Bochum, D-44780 Bochum, Germany\\
$^{(2)}$Institute of Radioengineering and Electronics of the Russian Academy%
\\
of Sciences, 103907 Moscow, Russia \\
$^{(3)}$L.D. Landau Institute for Theoretical Physics, 117940 Moscow, Russia }
\maketitle

\begin{abstract}
We calculate the dc Josephson current for two  superconductor-ferromagnet (S/F) bilayers separated by a
thin insulating film. It is demonstrated that the critical Josephson current 
$I_{c}$ in the junction strongly depends on the relative orientation of the
effective exchange field $h$ of the bilayers. We found that in the case of
an antiparallel orientation, $I_{c}$ increases at low temperatures with
increasing $h$ and at zero temperature has a singularity when $h$ equals the
superconducting gap $\Delta $. This striking behavior contrasts suppression
of the critical current by the magnetic moments aligned in parallel and is
an interesting new effect of the interplay between superconductors and
ferromagnets.
\end{abstract}

The possibility of  various  applications and the appearance of new
interesting physics  makes the  experimental and theoretical study of
ferromagnetic and superconducting-ferromagnetic hybrid structures    a popular topic. One of the properties that
has attracted in the last years a
lot of interest is a magnetoresistance due to the presence of the magnetic
order \cite{baibich,binasch,garcia1,garcia2}. In some structures the magnetoresistance can reach very large values. This effect has been termed ``giant magnetoresistance'' (GMR).
First discovered in magnetic multilayers \cite{baibich,binasch} where the
typical values of MR were of order of $10\%$, the GMR effect can be as large
as $200\%-300\%$ in $Ni-Ni$ or $Co-Co$ point contacts \cite{garcia1,garcia2}.

A typical device studied in such experiments consists of two separated
ferromagnets. One measures the resistivity for different  relative directions of the magnetization. The large values of
the MR is  due to an additional scattering of electrons at the
    boundary between adjacent layers (in the case of antiparallel orientation,
    an electron crossing this boundary goes  from one sub-band to another 
    and experiences a reflection from an effective  potential related to the different
    positions of the sub-bands) 

 If the normal metals of the reservoirs are replaced by superconductors,
another mechanism causes differing resistances
for the antiparallel and the parallel alignment of magnetization. This
mechanism is due to  Andreev reflection which occurs at the S/F interfaces, and
which implies a  zero spin current through them \cite{falko}. In the case of very
thin magnetic layers separating the superconducting reservoirs, the resistance
of the structure drops to zero and it becomes more appropriate to consider the  supercurrent (or Josephson
current). It was shown that if the exchange field $h$ in the magnetic layer
exceeds a certain value, the state energetically more favorable corresponds
not to a zero phase difference between the reservoirs (in the absence of an external current), but to a
phase difference of $\varphi = \pi$ (the so-called
$\pi$-junction)\cite{Buzdin}. The predicted $\pi$-state in a S/F/S Josephson
junction apparently was observed by Ryazanov {\em et al.} \cite{Ryaz}. The
critical current decreases with increasing exchange field $h$ in the magnetic
layer, changes sign and decays to zero while undergoing some oscillations. The
superconducting properties are not so strongly reduced if the magnetization
(i.e. the exchange field $h$) is not homogeneous \cite{Bul1,Berger}.

In this Letter we demonstrate that, in contrast to the common knowledge,
 the exchange field  can under certain conditions enhance the
Josephson critical current in a S/F-I-S/F tunnel junction rather than reduce
it (here I is an insulating layer). As a result, the
critical current $I_c$ may considerably exceed the critical current of the
Josephson junction  in the absence of the exchange field. The conditions are quite
simple: one needs low temperatures and  the antiparallel alignment of the
magnetization in the different parts of the superconductor. At the same
time, if the magnetization in the bilayers are parallel the critical current
is suppressed. This leads to a high sensitivity of the critical current to
the mutual alignment of the magnetic moments and, hence, to a possibility of
an experimental observation.

\begin{figure}
\epsfysize = 3.5cm
\vspace{0.2cm}
\centerline{\epsfbox{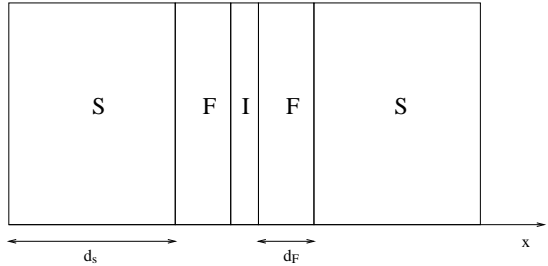
 }}
\vspace{0.2cm}
\caption{\label{Fig.1}The S/F-I-F/S system.}
\end{figure}

To be specific we consider a system consisting of two
superconductor-ferromagnet (S/F) bilayers (F here is a thin film) separated
by a thin insulating layer (see Fig.\ref{Fig.1}), i.e. the Josephson
S/F-I-F/S junction. This system can be
studied using quasiclassical equations \cite{Eilen,LOvchQuas,Usadel}
complemented with the boundary conditions \cite{Zaitsev,Kupr}. This approach allows one to describe the system
completely and was used to get the main results of the present paper.

However, the Josephson current and other thermodynamic quantities can be
derived in a considerably simpler way if the thicknesses of the layers $d_{S}$
and $d_F$
in
Fig.\ref{Fig.1} are   smaller than the superconducting coherence
    length $\xi_{S}\sim \sqrt{D/2\pi T_{c}}$ and the length of the condensate
    penetration into the ferromagnet $\xi_{F}\sim \sqrt{D/h}$,
    respectively. These conditions can be met experimentally.

Although, generally speaking, solutions for the superconducting order
parameter $\Delta $ of the quasiclassical equations depend on the
coordinates, the assumption  about the thickness allows one to write solutions that do
not have this dependence. In this limit, the influence of the ferromagnetic
layers on superconductivity is not local and is equivalent to inclusion of a
homogeneous exchange field with a reduced value. Of course, the other
physical quantities characterizing the superconductor should be modified,
too.
%%%%%%%%%%%%%%%%%%%%%%%%%%%%%%

%%%%%%%%%%%%%%%%%%%%%%%%%%%%%%%

Proceeding in this way, one comes to effective values of the superconducting
order parameter $\Delta _{eff}$, of the coupling constant $\lambda _{eff}$,
and of the magnetic moment $h_{eff}$ described by the following equations 
\[
\Delta _{eff}/\Delta =\lambda _{eff}/\lambda =\nu _{s}d_{s}\left( \nu
_{s}d_{s}+\nu _{f}d_{f}\right) ^{-1}, 
\]
\begin{equation}
h_{eff}/h=\nu _{f}d_{f}\left( \nu _{s}d_{s}+\nu _{f}d_{f}\right) ^{-1}
\label{a1}
\end{equation}
where $\nu _{s}$ and $\nu _{f}$ are the densities of states in the
superconductor and ferromagnet, respectively.

Assuming that the exchange field acts only on spin of electrons (which
implies that the magnetizations  are parallel to the interface) one can
write the Gor'kov equations for the S/F layers 
\[
\left( i \varepsilon _{n}+\xi -{\bf \sigma h}\right) \hat{G}%
_{\varepsilon }+\hat{\Delta}\hat{F}_{\varepsilon }^{+}=1
\]
\begin{equation}
\left( -i\varepsilon _{n}+\xi -{\bf \sigma h}\right) \hat{F}_{\varepsilon }+%
\hat{\Delta}\hat{G}_{\varepsilon }=0  \label{a2}
\end{equation}
where ${\bf \sigma }$ are Pauli matrices and $\xi =\varepsilon \left( {\bf p}%
\right) -\varepsilon _{F},$ $\varepsilon _{F}$ is the Fermi energy, $%
\varepsilon \left( {\bf p}\right) $ is the spectrum, $\varepsilon
_{n}=\left( 2n+1\right) \pi T$ are Matsubara frequencies, and $%
G_{\varepsilon }$ and $F_{\varepsilon }$ are normal and anomalous Green
functions. (We omit the subscript $eff$ \ in Eqs. (\ref{a2}) and below).
Eqs. (\ref{a2}) should be complemented by the self-consistency equation 
\begin{equation}
\Delta=\lambda T\sum_{\varepsilon }{\rm Tr}\hat{f}_{\varepsilon }
\label{a3}
\end{equation}
where trace ${\rm Tr}$ should be taken over the spin variables and 
\begin{equation}
\hat{f}_{\varepsilon }=\frac{1}{\pi }\int \hat{F}_{\varepsilon }d\xi 
\label{a4}
\end{equation}
Eqs. (\ref{a2}-\ref{a4}) may describe superconductors with a 
homogeneous exchange field as well. We neglect influence of the magnetic
moments on the orbital electron motion, which is definitely legitimate for
the thin ferromagnetic layers considered here. As soon as the S/F system is
described by Eqs. (\ref{a2}-\ref{a4})  the Josephson current $I_J$ can
    be expressed in terms of $\hat{f}$
\begin{equation}
I_{J}=\left( 2\pi T/eR\right){\rm Tr}\sum_{n}\hat{f}(h_1)\hat{f}(h_2)\sin \varphi  \label{a5}
\end{equation}
 where $R$ is the barrier resistance in the normal state. This formula
    can be easily obtained by using the standard tunneling Hamiltonian method or
    boundary conditions \cite{Zaitsev,Kupr}. $h_1$ and $h_2$ are the exchange
    fields to the left and to the right of the junction.

In the case of the conventional singlet superconducting pairing the matrix $%
\hat{\Delta}$ has the form $\hat{\Delta}=i\sigma _{y}\Delta $. Solving Eqs. (%
\ref{a2}) and using Eq. (\ref{a4}) we find easily for the function $\hat{f}%
_{\varepsilon }$%
\begin{equation}
\hat{f}_{\varepsilon }=\hat{\Delta}\left( \left( \varepsilon _{n}+i{\bf %
\sigma h}\right) ^{2}+\Delta ^{2}\right) ^{-1/2}  \label{a6}
\end{equation}

With Eq. (\ref{a6}) one can calculate the Josephson current $I_{J}$ for any
direction of the magnetic moments ${\bf h}_{1}$ and ${\bf h}_{2}$. The most
interesting are the cases of the parallel and antiparallel alignments of the
magnetic moments. In the both cases computation of the current $I_{J}$ in Eq.
(\ref{a5}) is very simple and we obtain for the parallel  configuration 
\begin{equation}
I_{J}^{\left( p\right) }\!\!=\frac{\Delta ^{2}\left( T\right) 4\pi T}{eR}\sum_{\varepsilon
}\frac{\varepsilon _{n}^{2}+\Delta ^{2}\left( T,h\right) -h^{2}}{\left(
\varepsilon _{n}^{2}+\Delta ^{2}\left( T,h\right) -h^{2}\right)
^{2}+4\varepsilon _{n}^{2}h^{2}},  \label{a7}
\end{equation}
whereas the Josephson current  $I^{\left( a\right) }$ for the antiparallel
configuration takes the form 
\begin{equation}
I_{J}^{\left( a\right) }\!\!=\frac{\Delta ^{2}\left( T\right) 4 \pi T}{eR}\sum_{\varepsilon
}\frac{1}{\sqrt{\left( \varepsilon _{n}^{2}+\Delta ^{2}\left( T,h\right)
-h^{2}\right) ^{2}+4\varepsilon _{n}^{2}h^{2}}}.  \label{a8}
\end{equation}
In Eqs. (\ref{a7}, \ref{a8}), $\Delta \left( T,h\right) $ is the
superconducting gap which depends on both the
temperature $T$ and the exchange field $h$ (for simplicity we assume that
the moduli of the exchange field are equal to each other). The value of the
superconducting order parameter $\Delta \left( T,h\right) $ is determined by
Eqs. (\ref{a3}, \ref{a6}) that can be reduced to the form 
\begin{equation}
1=\lambda \pi T\sum_{\varepsilon }
\mathop{\rm Re}
\frac{1}{\sqrt{\left( \varepsilon _{n}+ih\right) ^{2}+\Delta ^{2}\left(
T,h\right) }}\, .  \label{a9}
\end{equation}
Eqs. (\ref{a7}-\ref{a9}) solve completely the problem of calculation of the
Josephson energy and the critical current of the junction with the parallel
and antiparallel alignment of the magnetic moments and all new interesting
results of the present paper are described by these equations.

It is clear without further calculations that  the current $I_{c}^{\left( p\right) }$ of the parallel
configuration is always smaller than the  current  $I_{c}^{\left( a\right) }$
corresponding to the antiparallel one. So, rotating experimentally the magnetic moment of one of the S/F
bilayer one might considerably change the critical current.

Although this phenomenon is interesting on its own, Eq. (\ref{a8}) written
for the antiparallel alignment describes at low temperatures a much more
striking effect. In the limit $T\rightarrow 0$, the sums over the Matsubara
frequencies can be replaced by integrals and one obtains \cite{Fulde,LOv} 
\begin{equation}
\Delta \left( 0,h\right) =\left\{ 
\begin{array}{cc}
\Delta _{0}, & h<\Delta _{0} \\ 
0, & h>\Delta _{0}
\end{array}
\right.  \label{a11}
\end{equation}
where $\Delta _{0}$ is the BCS superconducting gap at $T=0$ in the absence
of the exchange field. There is another solution for $\Delta (h)<\Delta_0$ in
the interval $1/2<h<1$ \cite{LOv,Fulde}, but this solution is unstable.

Inserting Eq. (\ref{a11}) in Eq. (\ref{a8}) one can see that the Josephson
 critical current $I_{c}^{\left(a\right) }$ grows with increasing exchange field and even formally
logarithmically diverges when $h\rightarrow \Delta _{0}$%
\begin{equation}
I_{c}^{\left( a\right) }\left( h\rightarrow \Delta _{0}\right) \simeq
\frac{I_{c}\left( 0\right)}{\pi} \ln \left( \Delta _{0}/\omega _{0}\right)\, , 
\label{a12}
\end{equation}
where $I_{c}\left( 0\right) $ is the critical current in the absence of the
magnetic moment at $T=0$, and $\omega _{0}$ is a cutoff at low energies. 

At finite temperatures $\omega _{0}\sim T$ but, in principle, it should
remain finite also at $T=0$. The formal divergence seen in Eq. (\ref{a8})
can apparently be removed by considering  any damping in the excitation
    spectrum of the superconductors or higher orders in expansion in the
tunneling rate. 

The enhancement of the Josephson current by the presence of ordered magnetic
moments in superconductors, Eq. (\ref{a12}), is the main result of our paper
and is, to the best of our knowledge a novel effect. It occurs if the
magnetic moments are aligned {\it antiparallel}. In contrast, at finite temperature the Josephson
 critical current for a {\it parallel} alignment of the
magnetic moments are always smaller than the corresponding values without
the magnetic moments. At $T=0,$ the calculation of the integral over the
frequencies in Eq. (\ref{a7}) shows that $I_{c}^{(p)}$ does not depend on $h$,
coinciding with $I_{c}(0)$.

In principle, the dependence of the critical currents on the exchange field  can be more complicated due to a possibility of a transition to the
nonhomogeneous LOFF phase predicted by Larkin and Ovchinnikov (LO) \cite{LOv}
and Fulde and Ferrell \cite{Fulde} for the region $0.755\Delta _{0}<h<\Delta
_{0}$. Nevertheless, Eqs. (\ref{a7}-\ref{a12}) are applicable for $
h<0.755\Delta _{0}$, and a possible transition to the LOFF state would
manifest itself in a drop of the critical current.  Even for
    $h>0.755\Delta_{0}$ the predicted effect may survive because the state
with homogeneous $\Delta$ may exist as a metastable one.

\begin{figure}
\epsfysize = 5.5cm
\vspace{0.2cm}
\centerline{\epsfbox{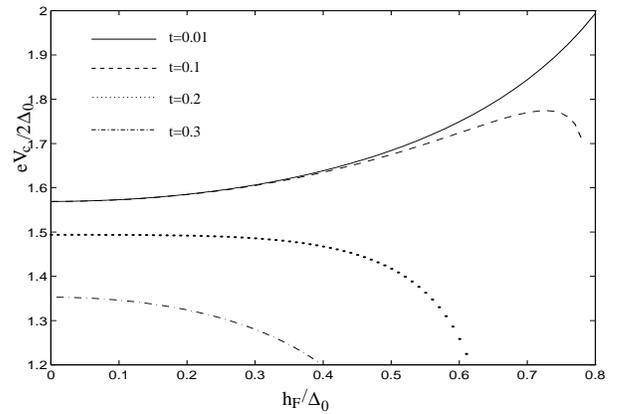
 }}
\vspace{0.2cm}
\caption{Dependence of the normalized critical current on $h$ for different
  temperatures in the case of an  antiparallel orientation. Here $eV_{c}=eRI_{c}$, $h_{F}$ is the
  effective exchange field, $t=T/\Delta_{0}$ and $\Delta_{0}$ is the
  superconducting order parameter at $T=0$ and $h=0$.\label{Fig.2}}
\end{figure}

\begin{figure}
\epsfysize = 5cm
\vspace{0.2cm}
\centerline{\epsfbox{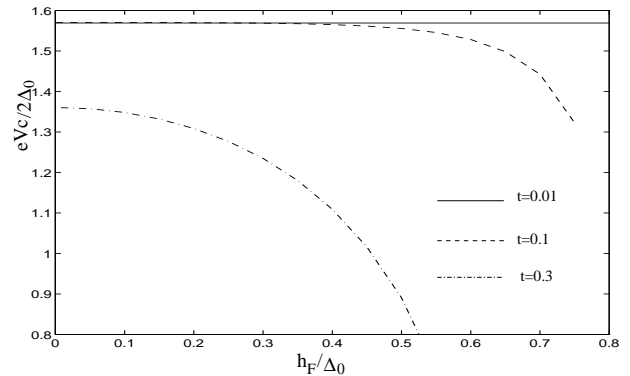
 }}
\vspace{0.2cm}
\caption{The same dependence as in Fig.2 in the case of a parallel orientation.\label{Fig.3}}
\end{figure}

The enhancement of the Josephson current occurs only at sufficiently low
temperatures. Near the transition temperature $T_{c\text{ }}$ and for small $h$ one obtains 
\[
I_{c}^{\left( a\right) }=\pi \left( eR \right)^{-1}\left( \Delta ^{2}/h
\right)\tanh \left(h/2T_{c}\right)\;,
\]
\[
I_{c}^{\left( p\right) }=\left( \pi /2\right)\left
  ( eR\right)^{-1}\left(\Delta^{2} /T_{c}\right)\cosh ^{-2}\left
  ( h/2T_{c}\right)\; ,
\]
\begin{equation}
I_{c}^{\left( a\right) }/I_{c}^{\left( p\right) }=\left( T_{c}/h\right)\sinh
\left( h/T_{c}\right)\, ,    \label{a14}
\end{equation}
where $\Delta=\Delta(T,h)$ is determined from Eq. (\ref{a9}).
The dependence of $T_c$ on $h$ is presented in Ref. \cite{sarma}. At arbitrary temperatures the dependence of the critical currents on the
exchange field $h$ can be obtained from Eqs. (\ref{a7}-\ref{a9}) only
numerically. The results are represented in Fig.\ref{Fig.2} for the antiparallel
configuration and in Fig.\ref{Fig.3} for the parallel one.   

%%%%%%%%%%%%%%%%%%%%%
%%%%%%%%%%%%%%%%%%%%%%%%%%%%
%%%%%%%%%%%%%%%%%%%%%%%%%%%%%%%%

If the angle $\alpha$ between the directions of the magnetization is
arbitrary the critical current $I_{c}^{\alpha }$ can be written in the form
\begin{equation}
I_{c}^{\alpha }=I_{c}^{\left( p\right) }\cos ^{2}\left( \alpha /2\right)
+I_{c}^{\left( a\right) }\sin ^{2}\left( \alpha /2\right)\, .   \label{a15}
\end{equation}
Eq. (\ref{a15}) shows that the singular part of the critical current is
always present and its contribution may reach $100\%$ at $\alpha =\pi$. 

All the conclusions presented above valid also for two magnetic
superconductors with uniformly oriented magnetization in each layer. Eqs. (\ref{a7}-\ref{a8}) could be obtained from formulae written in Ref. \cite{kulic} for magnetic superconductors with a spiral structure. However, the
effects found in our work were not discussed in Ref. \cite{kulic}. 

\begin{figure}
\epsfysize = 5.5cm
\vspace{0.2cm}
\centerline{\epsfbox{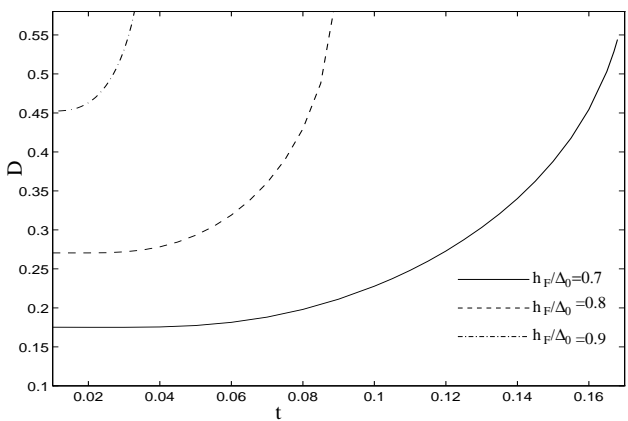
 }}
\vspace{0.2cm}
\caption{Temperature dependence of the coefficient $D$. Here $h_{F}$ is the
  effective exchange field and $t=T/\Delta_{0}$.\label{Fig.4}}
\end{figure}

Experimentally, it might be convenient to measure the coefficient $D$
\begin{equation}
D=\frac{I_{c}^{\left( a\right) }-I_{c}^{\left( p\right) }}{I_{c}^{\left(
p\right) }}  \label{a16}
\end{equation}
as a function of temperature. We draw in Fig.\ref{Fig.4} several curves
characterizing the temperature dependence $D\left( T\right)$ for different
$h$. One can change $h$ by varying the thickness of the magnetic layers. We see
that the coefficient $D$ can reach values of  the order of unity. We note that
at a given $h$ ($h>1/2$) a first order transition takes place when $T$ reaches a
certain critical value. In this case either $\Delta$ drops to
a smaller value or the normal state is realized.
If the S/F interface resistance per unit area $R_{S/F}$ exceeds the value
    $\rho_{F}d_{F}$ ($\rho_{F}$  is the specific resistance of the
    ferromagnet), the condensate functions experience a jump at the S/F
    interface and a sub gap
    $\epsilon_{sg}=\left(D\rho_{F}\right)_{F}/R_{S/F}d_{F}<\Delta$ arises in
    the ferromagnet \cite{mcmillan}. In this case a singularity appears when
    $h\rightarrow \epsilon_{sg}$.

%%%%%%%%%%%%%%%%%%%%%%%%%%%%
%%%%%%%%%%%%%%%%%%%%%%%%%%%%%

 All the results presented in this paper can be obtained by using the
    quasiclassical Green's function technique generalized for spin-dependent
    interaction. The details of the calculations will be presented
    elsewhere. It is important to mention that the enhancement of the
    Josephson current by the antiparallel alignment of the magnetic moments is
    obtained only for the singlet pairing.

 In conclusion, we have shown that in contrast to the common view, the
 presence of an    exchange field $h$ can increase the critical current $I_c$ in
    a Josephson tunnel junction S/F-I-F/S in the case of an antiparallel
    alignment of the magnetization in the ferromagnets.

We thank SFB 491 {\it Magnetische Heterostrukturen} for financial support.

\end{document}